Title: Photoinduced Electrification of Solids: II. Photovoltage Transients
Authors: M. Georgiev and O. Ivanov (Institute of Solid State Physics, Bulgarian
   Academy of Sciences, 72 Tsarigradsko Chaussee, 1784 Sofia, Bulgaria)
Comments: 18 pages, 2 figures, 1 table, pdf format
Journal-ref: To be submitted to J. Electrostatics

In relevance to the photoelectrification mechanisms proposed in Part I, we derive rate equations for the negative-U and STE photoelectrification modes and solve them under conditions close to experimental ones. Exact solutions are obtained for the case of a slow electron-hole recombination at negative-U sites. These solutions are compared with photovoltage versus time transients by "short-circuit currents" in photocharging experiments.

1. Introduction

In the preceding Part I to be referred to as I hereafter, we proposed a variety of mechanisms for the photoelectrification (photocharging) of poorly conducting solids effective in different spectral ranges of the excitation photons. In view of the specific conditions of pulsed illumination for studying the phenomenon, the experimentally measured quantity which provides the complete information on the particular electrification mechanism are the photovoltage transients.

A number of crystalline materials from insulators to compound semiconductors have been investigated, their photocharging voltages all falling within the µV range. Examples are shown in Figure 1 for (111) GaAs at 300 K.[1] These voltage forms obtained under pulsed "white" light from a Xe lamp are in concert with transients observed later under LED pulses within narrow spectral ranges at subgap or overgap energies. For investigating the photoresponse of a sample, Bergmann's condenser method has been employed in which the measuring capacitor forms between the upper face of the sample and a pressed metal-layer grounded electrode. Three kinds of a photosignal have been measured as triggered by the pulsed light in (a), namely: (b) a short-circuit photocurrent (PC) across a sample supplied with ohmic contacts on both its opposite faces; (c) the surface photo-voltage (SPV1) built up between the two opposite faces of the sample with only one ohmic contact, (d) the surface photovoltage (SPV2) across a sample with no ohmic contacts though pressed between two solid plates. The photoresponse in all the cases has been followed as the signal from the condenser has been fed to a high-input lock-in amplifier and then passed on to an analogue-digital analyzer. Care has been taken to account for voltage form distortions arising from the differentiating circuit and the finite RC constant (typically $10^{-3}$ s). Details can be complemented from the original publication.[1]

Other experiments were made more recently using photoexcitation by light emitting diodes (LED) within narrower spectral ranges at an overgap energy (655 nm (1.9 eV)) or at a subgap energy (950 nm (1.3 eV)). The measuring technique was as above, except for shining a shorter light pulse (3 ms) and for the photosignal passed from the lock-in amplifier to an oscilloscope for monitoring. The LED power was about 1 mW/cm$^2$. The subgap photovoltage proved qualitatively indistinguishable from SPV2 in Figure 1 (d).[1] However, the overgap voltage was likely the response of a free carrier photoelectret state composed of a polarization signal under light-on followed by a depolarization signal under light-off.[2] The chopping rate for the light pulses was chosen low (40 s$^{-1}$) so as to legitimize applying a single-pulse approach to the fundamental processes.

## 2. Kinetics of ion sputtering

The elementary steps of laser sputtering are subject to a math description in terms of basic rate processes. A self-consistent analysis yields specific waveforms produced by short-circuit currents in a photoelectrification experiment. These waveforms are signatures of the respective mechanism. We presently focus on solving the pertinent rate equations controlling the kinetics of elementary photocharging steps. Rate equations have been considered earlier as regards laser sputtering.[3,4] We now complement these equations so as to make them applicable to photocharging. The solutions are used to derive oscillographic wave forms in concert with the observed short-circuit voltages in photocharging. Examples of experimental waveforms are shown in Figure 1.

### 2.1. Photohole kinetics

The rate equations controlling the kinetics in a photoelectrification experiment are similar to the ones found appropriate for describing laser sputtering. We introduce p, $p_b$, and $p_d$ for the surface densities of uniholes, diholes and desorbed ions, respectively. N is the surface density of active bonds.

We believe it is instructive referring to a progenitor free unihole rate equation for a poorly conducting material:

$$d p / d t = G(t) - \gamma(p,T) \, m \, p \qquad (1)$$

in which $m$ is a carrier concentration whose nature is to be specified in each particular case. In what follows, we consider photohole kinetics controlled by negative-U ($m = p$) (A) and self-trapped exciton ($m = n$) formation (B).

### 2.2. Negative-U mode

Setting formally $m = p$ we construct a generic sequence of rate equations for negative-U dihole formation believed responsible for the photosputtering of metal ions from compound semiconductors:

$$d p / d t = G(t) - \gamma(p,T) \, p^2 \qquad (2)$$

$$d p_b / d t = \gamma(p,T) \, p^2 + (p_d / \tau_d - p_b / \tau_b) - p_b / \tau_r \qquad (3)$$

$$d p_d / d t = p_b / \tau_b - p_d / \tau_d \qquad (4)$$

Here G stands for the electron-hole generation rate, $\tau_b$ is the dihole lifetime against desorption and $\tau_d$ is the desorbed ion lifetime against readsorption. We set

$$G(p, p_b, t) = \eta\, c_\phi\, I(t)\, (N - 2p - p_b + p_d) \qquad (5)$$

for G(t). I(t) is the excitation light intensity, $c_\phi$ is the photon absorption cross section, and $\eta$ is the electron-hole generation quantum yield.

The parameter is the dihole trapping coefficient $\gamma(p,T)$ derived earlier:[3]

$$\gamma(p,T) = [\gamma_b(T)^{-1} + \gamma_d(p,T)^{-1}]^{-1} \qquad (6)$$

where the components read

$$\gamma_d(p,T) = (4\pi e \mu_h / \kappa)\, \exp(-\alpha[1/r_0 - 1/r_p]), \qquad (7)$$

$$\gamma_b(T) = \hbar^{-1}(N/W)\sqrt{\pi}\, (E_{LR} k_B T)\, (k_B T / \hbar\omega)\, \sinh(\hbar\omega / 2k_B T) \times$$

$$\{\sqrt{[(W+U)^2 / E_{LR} k_B T]}\, \exp(-\sqrt{[(W+U)^2 / E_{LR} k_B T]}) +$$

$$?\sqrt{\pi}\, [1 - \mathrm{erf}([W+U]^2 / E_{LR} k_B T)]\} \qquad (8)$$

with $r_p = \sqrt{(\kappa k_B T / 8\pi e^2 p)}$ (Debye's screening radius), $\alpha = e^2/\kappa k_B T$. Here $r_0$ is the polaron radius, $\mu_h$ is the hole mobility, $\kappa$ is the dielectric constant, W is the free hole bandwidth, $E_{LR}$ is the lattice relaxation energy, $U = U_0 - 4E_{LR}$ is the "negative-U energy" where $U_0 = (e^2 / \kappa r_0)\, \exp(-r_0 / r_p)$ is the screened Coulomb repulsion barrier. Equation (2) implies that the diffusion-limited trapping coefficient $\gamma_d$ is only competitive when Debye's radius $r_p$ is comparable to the polaron radius $r_0$ which comes at high enough photocarrier concentrations, *Itoh-Nakayama's negative-U barrier collapse.* Equation (8) is derived applying the reaction rate theory to a screened Coulomb potential: $U(r) = (e^2 / \kappa r)\, \exp(-r / r_p)$.

It should be stressed that equation (7) seems essential for reproducing theoretically the details of the basic experimental dependencies. Indeed, the observed sputtering yield *vs.* laser fluence in Figure 2-I has been confirmed through solving numerically for the nonlinear rate equations only after the full array of equations (6) through (8) for the hole trapping coefficient $\gamma$ have been accounted for.[4] Yet, a density independent $\gamma$ has been used for some analytical purposes and found feasible.[3]

The electron-hole recombination rate R(*n,p*) is considered sizeable at negative-U hole sites only and is assumed low elsewhere because of the spatial segregation of photoelectrons from photoholes, due to their separate capture at dangling and occupied surface bonds, respectively. The form R(*n,p*) = $p_b / \tau_r$ is adopted here with the understanding that the recombination time $\tau_r$ may be constant in specific cases only. One is the case of an *n*-type semiconducting material, such as the Cr-compensated *n*-GaAs, the majority carrier concentration *n* assumed constant leading to a constant $\tau_r = 1 / \gamma_r n$.

The negative-sign dihole contribution to G accounts for the reduction in the occupied bond density following the captute of a dihole which causes that bond to break and an ion to desorb. We believe desorption occurs almost automatically (desorption time ~ 1 ps) following the breaking of a bond. For the sake of convenience, we categorize the desorbed ions as outgoing ($p_b$) and incoming ($p_d$). The average lifetime of an outgoing ion before recharging at the outer electrode is $\tau_b$. It will give a positive flow out of the surface. On going back to the surface during a dark period, an incoming ion will give a dihole back to the lattice which will recombine with a dangling bond dielectron to restore the initial state of the crystal prior to the illumination. A finite average lifetime $\tau_d$ will be attributed to a re-adsorbed ion before returning to the surface to give back its hole charge. It will produce a negative flow towards the surface.

Upon substituting G into the rate equations, a Riccati type ordinary differential equation results for p(t).[5] The complete differential equation for p(t) is nonlinear and has been solved only numerically. An iterative solution has been found for the light period ($I \neq 0$) assuming infinite dihole and desorbed ion lifetimes $\tau_b = \tau_d = \infty$.[4] Assuming $\gamma(p,T) \equiv \gamma(T)$, the trapping parameter calculated at constant carrier density $p$, the complete differential equation has been linearized and analyzed analytically. Numerical and analytic conclusions drawn for $I \neq 0$ have been found in concert.

We remind that in accordance with eq.(7):

$$\gamma_d(p,T) = \gamma_d(0,T) \exp(\alpha/r_p) = \gamma_d(0,T) \exp(b\sqrt{p})$$

$$\gamma_d(0,T) = (4\pi e \mu_h /\kappa) \exp(-\alpha/r_0)$$

where $\alpha/r_p = e^3 \sqrt{8\pi} (\kappa k_B T)^{-3/2} \sqrt{p} = b\sqrt{p}$, $b = e^3 \sqrt{8\pi} (\kappa k_B T)^{-3/2}$. For an estimate we set $\kappa = 5$ to obtain $b = 5.89 \times 10^{-9}$; for $p = 10^{11}$ cm$^{-3}$ this implies $b\sqrt{p} \sim 0.002 \ll 1$. Inserting into eq. (2) we get

$$dp/dt = G(t) - \gamma(0,T) \exp(b\sqrt{p}) p^2 \qquad (9)$$

Separating the variables we obtain writing $\gamma_0 \equiv \gamma_d(0,T)$:

$$t - t_i = {}_{pi}\!\int^p dp / [G(t) - \gamma_0 \exp(b\sqrt{p}) p^2] \qquad (10)$$

This is the complete equation of the free-hole density accounting for the Itoh-Nakayama proposal of eq. (7). The complete equation being mathematically more complex, the linearized theory at $b = 0$ may fully suffice for revealing the essential physics, as shown below.

Riccati's rate equations are easier to deal with at $N \gg 2p, p_b, p_d$ which implies $G(t) = \eta c_\phi N I(t)$. Inasmuch as the light intensity is a step function of time: $I(t) = I_p$ for $t < t_p$ and $I(t) = 0$ for $t > t_p$ where $t_p$ is the pulse time, the kinetic equation for p(t) is one with separable variables. Solving for eq. (2), we get for light-on ($t < t_p$, $G \neq 0$) under $p_i = 0$

at $t_i = 0$:

$$p(t)_{on} = \sqrt{(G/\gamma)} \tanh(\sqrt{(G\gamma)}\, t) \tag{11}$$

For light-off ($t > t_p$, $G = 0$) under $p = p_p$ at $t = t_p$ we get:

$$p(t)_{off} = p_p / [1 + \gamma\, p_p(t - t_p)] \tag{12}$$

$p_p$ tends to the saturated free-hole density

$$p_s = \sqrt{(G/\gamma)} \tag{13}$$

for $t_p \gg \sqrt{(G\gamma)}^{-1}$.

Solving for the rate equations will be described briefly for the negative-U mode first. It goes along the following lines: Summing up the second and the third equations we get

$$dp_b/dt + p_b/\tau_r = \gamma p^2 - dp_d/dt \tag{14}$$

which is readily integrated to give:

$$p_b(t) = -p_d(t) + [p_b(t_i) + p_d(t_i)] \exp(-(t-t_i)/\tau_r) +$$

$$\exp(-t/\tau_r) \int_{t_i}^{t} \{\gamma\, [p(t')]^2 + p_d(t')/\tau_r\} \exp(t'/\tau_r)\, dt' \tag{15}$$

if regarded as a differential equation for $p_b$, or it turns into an equivalent integral equation if integrated term by term:

$$p_b(t) + \int_{t_i}^{t} [p_b(t')/\tau_r]\, dt' =$$

$$[p_b(t_i) + p_d(t_i)] - p_d(t) + \int_{t_i}^{t} \gamma\, [p(t')]^2\, dt' \tag{16}$$

The inclusion of a finite recombination time $\tau_r$ drastically affects the immediate analytic conclusions. There are two extreme cases of electron-hole recombination, making also use of eq. (3):

(i)   slow in which

$$\int_{t_i}^{t} [p_b(t')/\tau_r]\, dt' \ll p_b(t) \text{ or } 2 p_b/\tau_r \ll \gamma p^2 + p_d/\tau_d - p_b/\tau_b$$

(ii)   fast in which

$$\int_{t_i}^{t} [p_b(t')/\tau_r]\, dt' \gg p_b(t) \text{ or } 2 p_b/\tau_r \gg \gamma p^2 + p_d/\tau_d - p_b/\tau_b$$

### 2.2.1. Slow electron-hole recombination

For an analytic study, we first take up case (i) of an infinitely large $\tau_r$. We exclude $p_b(t)$ by inserting eq. (16) into eq. (3) and solving at $\tau_r = \infty$ to get:

$$p_b(t) = -p_d(t) + [p_b(t_i) + p_d(t_i)] + J(t) \tag{17}$$

$$p_d(t) = \exp(-t[1/\tau_d + 1/\tau_b]) \times \{ p_d(t_i) \exp(t_i[1/\tau_d + 1/\tau_b]) +$$

$$\int_{t_i}^{t} \{[p_b(t_i) + p_d(t_i)] + J(t')\} \exp(t'[1/\tau_d + 1/\tau_b]) \, dt'/\tau_b \} \tag{18}$$

In both cases the main mathematics is in evaluating the integrals

$$J(t) = \int_{t_i}^{t} \gamma [p(t')]^2 \, dt' \tag{19}$$

which enter as source functions for the ionic component. To find the relevant solutions one should insert the pertinent initial conditions at $t = t_i$ for either light-on or light-off and perform the integrations.

For light-on we derive:

$$J(t)_{on} = G\,t \tag{20}$$

For light-off we find straightforwardly:

$$J(t)_{off} = p_p \{1 - [1 + \gamma\, p_p(t - t_p)]^{-1}\} \tag{21}$$

Inserting J(t), we get for light-on ($t_i = 0$, $p_i = ... = 0$):

$$p_d(t)_{on} = G(1 + \tau_b/\tau_d)^{-1} \{ t - (1/\tau_d + 1/\tau_b)^{-1} \times$$

$$[1 - \exp(-t[1/\tau_d + 1/\tau_b])] \} \tag{22}$$

while for light-off ($t_i = t_p$, $p_i = p_p$, etc.) we obtain using eq. (17):

$$p_d(t)_{off} = [G\,t_p + p_p](1 + \tau_b/\tau_d)^{-1}[1 - \exp(-[1/\tau_d + 1/\tau_b](t - t_p))] -$$

$$(\gamma\tau_b)^{-1} \exp([t_p - 1/\gamma p_p][1/\tau_d + 1/\tau_b]) \exp(-t[1/\tau_d + 1/\tau_b]) \times$$

$$\{\text{Ei}([(1/\tau_d + 1/\tau_b)/\gamma p_p][1 + \gamma p_p(t - t_p)]) - \text{Ei}([(1/\tau_d + 1/\tau_b)/\gamma p_p])\}$$

$$+ p_d(t_p) \exp(-(t - t_p)[1/\tau_b + 1/\tau_d]) \tag{23}$$

Ei(x) is the exponential integral function $\text{Ei}(x) = -\int_{-x}^{\infty} (e^{-t}/t)\,dt$.[6] We arrive at

$$p_b(t)_{on} = -p_d(t)_{on} + G\,t \tag{24}$$

$$p_b(t)_{off} = -p_d(t)_{off} + G\,t_p + p_p\{1 - [1 + \gamma p_p(t - t_p)]^{-1}\} \tag{25}$$

### 2.2.2. Fast electron-hole recombination

It should be noted that the solutions by (i) under $\tau_r = \infty$ are exact. To solve under a finite $\tau_r$ we have to make a simplifying assumption as in (ii) in view of the math complexity of the complete problem.

In case (ii) of predominating electron-hole recombination, we get from eq. (14) neglecting $d\,p_b/d\,t$:

$$p_b(t)/\tau_r = -\,d\,p_d(t)/dt + \gamma\,[\,p(t)\,]^2 \qquad (26)$$

Inserting $d\,p_d/d\,t$ from eq. (4) we find

$$p_b(t) = (1/\tau_r + 1/\tau_b)^{-1}\,\{\,p_d/\tau_d + \gamma\,[\,p(t)\,]^2\,\} \qquad (27)$$

which will be inserted back into eq. (4) with the result

$$d\,p_d/d\,t + p_d/\tau = (1+\tau_b/\tau_r)^{-1}\,\gamma\,[\,p(t)\,]^2 \qquad (28)$$

This is readily solved to give

$$p_d(t) = \exp(-t/\tau)\,\{\,p_d(t_i)\,\exp(t_i/\tau) + $$
$$(1+\tau_b/\tau_r)^{-1}\,{}_{t_i}\!\int^t \gamma\,[\,p(t')\,]^2\,\exp(t'/\tau)\,d\,t'\,\} \qquad (29)$$

with $\tau = \tau_d\,\{\,\tau_d\,(1+\tau_b/\tau_r)\,/\,[\tau_d\,(1+\tau_b/\tau_r) - 1]\,\}$.

The integral in eq. (29) should be evaluated separately for light-on and light-off. For the light period we get:

$${}_{t_i}\!\int^t \gamma\,[p(t')_{on}]^2\,\exp(t'/\tau)dt' = \sqrt{(G\,\gamma)}\,\tau\,[1 - \exp(t/\tau)] + $$
$$\sqrt{(G/\gamma)}\,\{-\tanh(\sqrt{(G\,\gamma)}\,t)\,\exp(t/\tau) + $$
$$(1/\tau)\,{}_0\!\int^t \tanh(\sqrt{(G\,\gamma)}\,t')\,\exp(t'/\tau)\,d\,t'\,\} \qquad (30)$$

The remaining integral is incorporated in

$$I(t) = \sqrt{(G/\gamma)}(1/\tau)\,\exp(-t/\tau) \times {}_0\!\int^t \tanh(\sqrt{(G\,\gamma)}\,t')\,\exp(t'/\tau)\,dt'.$$

This integral diverges at large t. To evaluate I(t), we apply l'Hospital's rule which gives

$$I(t) = \sqrt{(G/\gamma)}\,\tanh(\sqrt{(G\,\gamma)}\,t).$$

We thus get finally for $p_d(t)_{on}$ at $p_d(0)_{on} = 0$:

$$p_d(t)_{on} = (1 + \tau_b / \tau_r)^{-1} G \tau [1 - \exp(-t/\tau)] \tag{31}$$

while $p_b(t)_{on}$ obtains from eq. (27):

$$p_b(t)_{on} = (1/\tau_r + 1/\tau_b)^{-1} \{ p_d(t)_{on} / \tau_d + \gamma [p(t)_{on}]^2 \} \tag{32}$$

For the dark period

$$\int_{t_i}^{t} \gamma [p(t')_{off}]^2 \exp(t'/\tau) dt' = p_p \exp(1 - \gamma p_p t_p) \times$$

$$\{ \exp(1/\gamma p_p \tau) - (1/[1 + \gamma p_p(t - t_p)]) \exp([1 + \gamma p_p(t - t_p)]/\gamma p_p \tau)$$

$$+ (1/\gamma p_p \tau)[\operatorname{Ei}([1 + \gamma p_p(t - t_p)]/\gamma p_p \tau) - \operatorname{Ei}(1/\gamma p_p \tau)] \} \tag{33}$$

From eq. (28) we get for $p_d(t)_{off}$:

$$p_d(t)_{off} = p_d(t_p) \exp(-(t - t_p)/\tau) +$$

$$(1 + \tau_b / \tau_r)^{-1} p_p \exp(1 - \gamma p_p t_p) \exp(-t/\tau) \times$$

$$\{ \exp(1/\gamma p_p \tau) - (1/[1 + \gamma p_p(t - t_p)]) \exp([1 + \gamma p_p(t - t_p)]/\gamma p_p \tau)$$

$$+ (1/\gamma p_p \tau)[\operatorname{Ei}([1 + \gamma p_p(t - t_p)]/\gamma p_p \tau) - \operatorname{Ei}(1/\gamma p_p \tau)] \} \tag{34}$$

whereas $p_b(t)_{off}$ is again obtainable from eq. (27):

$$p_b(t)_{off} = (1/\tau_r + 1/\tau_b)^{-1} \{ p_d(t)_{off} / \tau_d + \gamma [p(t)_{off}]^2 \} \tag{35}$$

### 2.3. Self-trapped exciton mode

Next we set $m = n = $ const in which $n$ is the electron concentration, e.g. the majority carrier concentration compensated by Cr impurities in high-ohmic GaAs, to define the corresponding sequence of generic rate equations for self-trapped exciton formation. We believe the latter is responsible for the laser sputtering of metalloid ions from binary compounds, such as the Cr-compensated GaAs. We again introduce p, $p_b$ and $p_d$ for the surface density of uniholes, self-trapped excitons and desorbed ions, respectively. The pertinent unihole equation is:

$$dp/dt = G(t) - p/\tau_s \tag{36}$$

where $\tau_s = [\gamma(p,T) n]^{-1}$ is the average formation time of a self-trapped exciton out of the sea of free excitons.[7,8] The solution at $G = $ const reads

$$p(t)_{on} = G \tau_s [1 - \exp(-t/\tau_s)], \; t \leq t_p \tag{37}$$

$$p(t)_{off} = p(t_p) \exp(-(t-t_p)/\tau_s), \; t \geq t_p \tag{38}$$

Inserting into

$$J(t) = {}_{ti}\int^t [p(t')/\tau_s] \, dt' \tag{39}$$

we get

$$J(t)_{on} = G\{t - \tau_s[1 - \exp(-t/\tau_s)]\}, \; {}_{ti\,=\,0} \tag{40}$$

$$J(t)_{off} = p(t_p)[1 - \exp(-(t-t_p)/\tau_s)]\}, \; {}_{ti\,=\,tp} \tag{41}$$

to be used as a source function for the ionic components, as before.

The integral equation (16) is reformulated to read

$$p_b(t) + {}_{ti}\int^t [p_b(t')/\tau_r] \, dt' =$$

$$[p_b(t_i) + p_d(t_i)] - p_d(t) + {}_{ti}\int^t [p(t')/\tau_s] \, dt' \tag{42}$$

We specify two cases of negligible (i) and strong (ii) electron-hole recombination defined, respectively, by

(i) $p_b(t) \gg {}_{ti}\int^t [p_b(t')/\tau_r] \, dt'$ or $2 p_b/\tau_r \ll p/\tau_s + p_d/\tau_d - p_b/\tau_b$

(ii) $p_b(t) \ll {}_{ti}\int^t [p_b(t')/\tau_r] \, dt'$ or $2 p_b/\tau_r \gg p/\tau_s + p_d/\tau_d - p_b/\tau_b$

Here $\tau_r$ is the average electron-hole recombination time at the exciton.

### 2.3.1. Slow electron-hole recombination

As before we set $\tau_r = \infty$ to discard the electron-hole recombination contribution if it is too slow to affect the hole kinetics within the observation time range. The ionic densities formally satisfy the same rate equations as above which yield the solutions (17) and (18).

Combining we get for light-on under $t_i = 0$, $p(t_i) = p_b(t_i) = p_d(t_i) = 0$:

$$p_d(t)_{on} = G[1+\tau_b/\tau_d]^{-1}\{t - [1/\tau_d+1/\tau_b]^{-1}[1 - \exp(-t[1/\tau_d+1/\tau_b])]\} -$$

$$G(\tau_s/\tau_b)\{[1/\tau_d+1/\tau_b]^{-1}[1 - \exp(-t[1/\tau_d+1/\tau_b])] -$$

$$[1/\tau_d+1/\tau_b-1/\tau_s]^{-1}[\exp(-t/\tau_s) - \exp(-t[1/\tau_d + 1/\tau_b])]\} \tag{43}$$

and for light-off under $t_i = t_p$, $p(t_p)_{off} = p(t_p)_{on}$, $p_b(t_p)_{off} = p_b(t_p)_{on}$, $p_d(t_p)_{off} = p_d(t_p)_{on}$:

$$p_d(t)_{off} = [p_b(t_p) + p_d(t_p)][1 + \tau_b/\tau_d]^{-1}\{1 - \exp(-(t-t_p)[1/\tau_d + 1/\tau_b])\} +$$

$$p(t_p)\{\exp(-t_p/\tau_s)[1 + \tau_b/\tau_d]^{-1}\{1 - \exp(-(t-t_p)[1/\tau_d + 1/\tau_b])\} -$$

$$[1 + \tau_b/\tau_d - \tau_b/\tau_s]^{-1}\{\exp(-t/\tau_s) - \exp(-(t-t_p)[1/\tau_d + 1/\tau_b])\exp(-t_p/\tau_s)\}\}$$

$$+ p_d(t_p)\exp(-(t-t_p)[1/\tau_b + 1/\tau_d]) \tag{44}$$

Once the ion density $p_d(t)$ has been derived, the self-trapped exciton concentration $p_b(t)$ obtains from

$$p_b(t)_{on} = -p_d(t)_{on} + G\{t - \tau_s[1 - \exp(-t/\tau_s)]\} \tag{45}$$

$$p_b(t)_{off} = -p_d(t)_{off} + [p_b(t_p)_{on} + p_d(t_p)_{on}] +$$

$$p(t_p)[1 - \exp(-(t-t_p)/\tau_s)]\} \tag{46}$$

### 2.3.2. Fast electron-hole recombination

The case (ii) integral equation (41) is differentiated on its both sides resulting in

$$p_b(t) = (1/\tau_r + 1/\tau_b)^{-1}[p_d(t)/\tau_d + p(t)/\tau_s] \tag{47}$$

using eq. (7). $p(t)$ is defined by eq. (36) and (37) for -on and -off, respectively. Inserting $p_b$ into eq. (7) we get

$$dp_d/dt + p_d/\tau = [(1 + \tau_b/\tau_r)\tau_s]^{-1} p(t)$$

$$\tau^{-1} = \tau_d^{-1}[1 - (1 + \tau_b/\tau_r)^{-1}] \tag{48}$$

which is solved to give

$$p_d(t) = p_d(t_i) \exp(-(t - t_i)/\tau) +$$

$$\exp(-t/\tau)[(1 + \tau_b/\tau_r)\tau_s]^{-1} \int_{t_i}^{t} p(t') \exp(t'/\tau) dt' \tag{49}$$

Using eq. (36) and (37) we find:

$$p_d(t)_{on} = G(1 + \tau_b/\tau_r)^{-1}\{\tau[1 - \exp(-t/\tau)] -$$

$$(1/\tau + 1/\tau_s)^{-1}[\exp(-t/\tau_s) - \exp(-t/\tau)]\} \tag{50}$$

$$p_d(t)_{off} = p(t_p)[(1 + \tau_b/\tau_r)\tau_s]^{-1}[1/\tau_s + 1/\tau]^{-1}\{\exp(-(t-t_p)/\tau_s) -$$

$$\exp(-t/\tau)\} \tag{51}$$

We finally get from eq. (47):

$$p_b(t)_{on} = (1/\tau_r + 1/\tau_b)^{-1} [p_d(t)_{on}/\tau_d + p(t)_{on}/\tau_s] \qquad (52)$$

$$p_b(t)_{off} = (1/\tau_r + 1/\tau_b)^{-1} [p_d(t)_{off}/\tau_d + p(t)_{off}/\tau_s] \qquad (53)$$

Solving for the concentrations $p_b$ and $p_d$ is seen to be more or less elementary.

## 2.4. Photoinduced conductivity

The solutions derived hitherto are summarized in Table I. Ei($x$) therein is the exponential integral function Ei $(x) = -\int_{-x}^{\infty} (e^{-t}/t) dt$.[6] For both sputtering modes, negative-U and STE, the "short-circuit current conductivity" will be

$$\sigma(t) = e\,\mu_p\,p(t) + 2e\,\mu_b\,p_b(t) - 2e\,\mu_d\,p_d(t) \qquad (54)$$

where for tentative purposes the carrier mobilities are ascribed the same numerical values $\mu_p = \mu_b = \mu_d = 1$ cm$^2$V$^{-1}$s$^{-1}$. This order of magnitude is regarded a compromise for ionic, hole and bipolaron mobilities in a binary compound. The electric charge of an incoming ion is assumed to be the same as the one of an outgoing "*Anderson bipolaron*", if complete relaxation of the charge is to take place during the dark period.

We apply the obtained solutions for the photoinduced conductivities (see eq. (54)) to comparing with the experimental photocurrent and voltage transients observed under pulsed illumination across the GaAs bandgap at 871 nm (1.42 eV at 300 K).[9] Physically, the following scenario appears plausible: On photoexcitation at an overgap energy (LED at 655 nm (1.9 eV)) fast photoelectrons will be moving to surface dangling bonds. Trapped electrons will charge the surface negatively, creating a *photoelectret state by free carriers*,[2] giving rise to an electric field which will drag photoholes towards it. A slower growing positive hole response may appear superimposed on the voltage signal during the light pulse. At light-off an electronic depolarization spike may be observed eventually as the hole component decays more slowly. Photoexcitation at a subgap energy (LED at 950 nm (1.3 eV)) will give rise to dilute photoholes possibly originating from subsurface states. Now the low voltage formed during the light pulse may be due to desorbing ions mainly (positive signal) which decays in the dark as readsorbing ions move back to the surface (negative signal). Finally, the photovoltage relaxes in the dark due to electron-hole recombination.

As far as the voltage form is concerned, more appealing are the photoinduced transients reproduced in Figure 1 (a) through (d).[1] We notice fast components on both light-on and light-off in PC which do not seem present in SPV1. The SPV2 case is perhaps most appropriate for checking our sputtering proposal because of the lack of any adsorbed ohmic layers. Indeed, the photovoltage is seen to first rise, pass through a peak and then decline, all during the light pulse. During the subsequent dark period the voltage declines further and even becomes negative, drops to a low and then gradually rises less negative before relaxing completely. The main qualitative features of the photovoltage transients are seen reproduced in Figure 2 by the solutions for a slow electron-hole recombination, including the peak under light-on, the negative dip in signal and the voltage vanishing at longer times under light-off. In partucular, the negative voltage signal comes from the predominating contribution of reabsorbing ions at the later relaxation stages. The signal relaxes finally due to electron-hole recombination.

### 3. Discussion

Elsewhere, short-circuit currents seem to have been observed in semiconductors under pulsed illumination which are ascribed to a variety of mechanisms, such as the photovoltaic effect and others. Some crystals have generated photovoltages of magnitudes on the μV scale comparable to ours, such as a short-circuit current from ferroelectric $Ba_{0.25}Sr_{0.75}Nb_2O_6$.[2] However differences are clearly manifested on comparing with the photocharge transients in Figure 1 taken on highly ohmic *n*-GaAs. A fast component is attributed to a photovoltaic effect. The transiet is a pyroelectric current, due to laser heating. Its reverse polarity is understandable as the current flows in the internal photovoltaic field. Remarkably, the pyroelectric component disappears, while the photo-emf drops in magnitude on going to the paraphase.

The solutions for the pulsed photoconductivity as extracted in Table I are shown in Figure 2, (a), (c), (e) for light-on and (b), (d), (f) for light-off. These should be compared with the respective SPV2 transients in Figure 1 (d), respectively. We see the conductivities generally in concert with the observed voltage transients.[1]

The role of reverse processes of ion desorption and readsorption back to the surface along with the increased detection sensitivity comes to explain just why photocharging has only been observed under pulsed light rather than under incandescent lamp illumination. This is confirmed by the conductivity peak in Figure 1 (e) followed by a declining photovoltage in (f) down to negative values all during the light pulse.

### Acknowledgement

We thank Dr. L. Konstantinov (Sofia) for his interest and support.

References

[1] I. Davydov, O. Ivanov, D. Svircov, G. Georgiev, A. Odrinsky, and V. Pustovoit, Spectroscopic Lett. 27, 1281-1288 (1994).

[2] V.M. Fridkin, *Ferroelectrics - Semiconductors* (Nauka, Moscow, 1976) (in Russian).

[3] M. Georgiev and J. Singh, Appl. Phys. A **55**, 170-175 (1992).

[4] M. Georgiev, L. Mihailov and J. Singh in: Proc. VIII International School on Condensed Matter Physics, Varna (John Wiley, New York, 1994), p. 507-510.

[5] E. Kamke, *Differential Gleichungen. I. Gewöhnliche Differentialgleichungen* (Leipzig, 1959). Russian translation: (Nauka, Moscow, 1971).

[6] I.S. Gradstein and I.M. Ryzhik, Table of Integrals, Series, and Products, A. Jeffrey, ed. (Academic, New York, 1965), p. 925.

[7] *Defect Processes Induced By Electronic Excitation In Insulators*, N. Itoh, ed. (World Scientific, Singapore, 1989); K, Tanimura, *ibid*., p. 178; N. Itoh, *ibid*., p. 253.

[8] N. Itoh, Nucl. Instrum. & Methods in Phys. Res. **B27**, 155-166 (1987).

[9] S.M. Sze, Physics of Semiconductor Devices, 2nd ed. (Wiley, New York, 1981), p. 849.

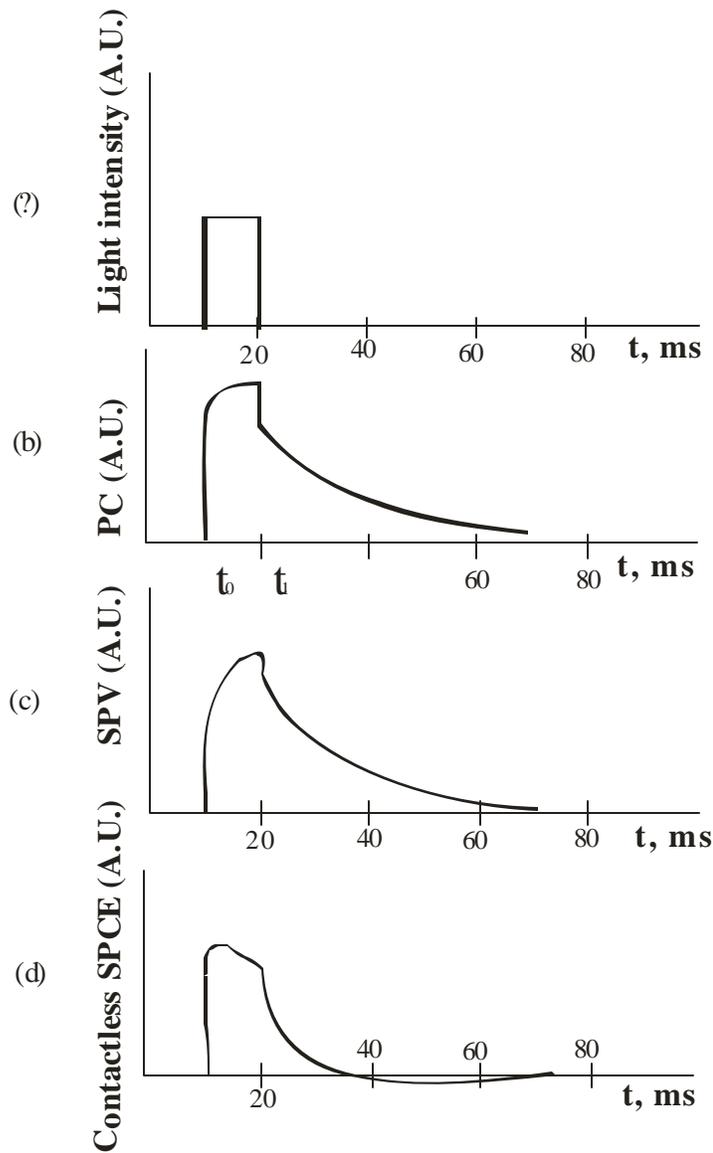

Figure 1. Induced voltage transients on photocharging a (111) GaAs surface. Three kinds of a photosignal have been monitored under the rectangular light pulses in (a), namely: (b) a short-circuit photocurrent (PC) across a sample supplied with ohmic contacts on both opposite faces; (c) the surface photovoltage (SPV1) built up between the opposite faces of a sample with only one ohmic contact, (d) the surface photovoltage (SPV2) across a sample with no ohmic contacts though pressed between two solid plates. The excitation light from a 500 W Xe lamp was chopped by a disk modulator at 8 c/s repetition rate. The photosignal from the sample-containing condenser was fed to a lock-in amplifier and then passed onto an analogue-digital converter. See reference [1] for details.

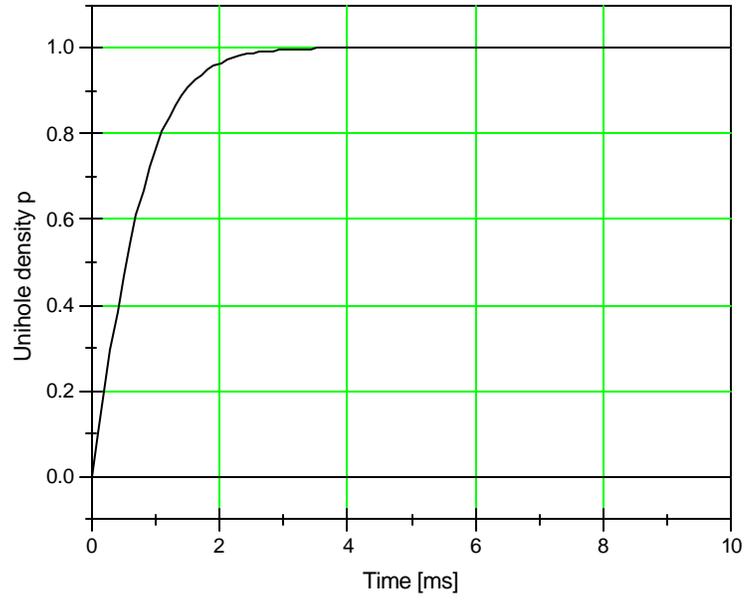

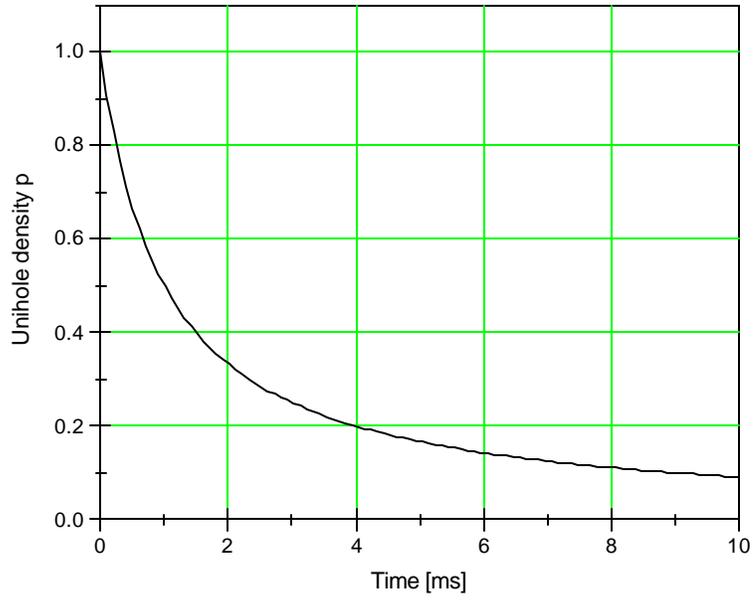

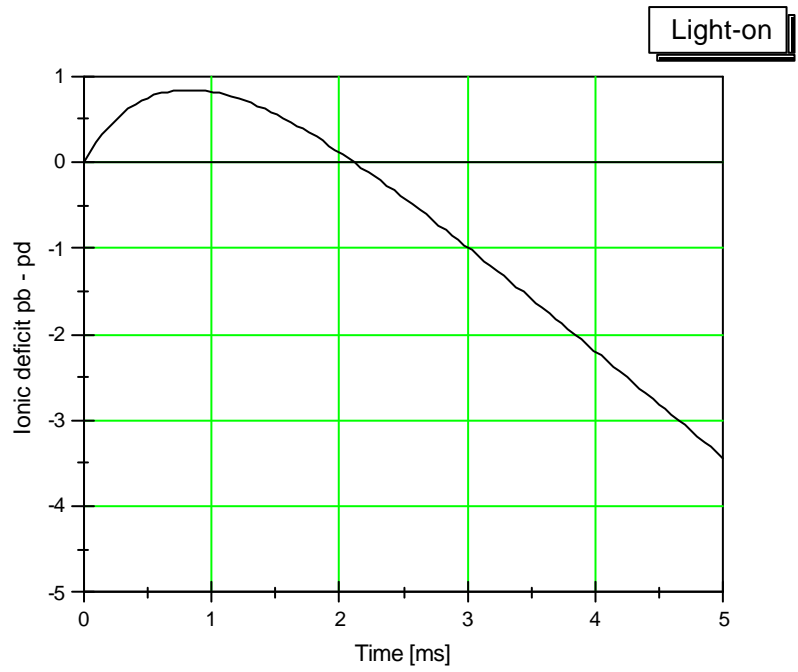

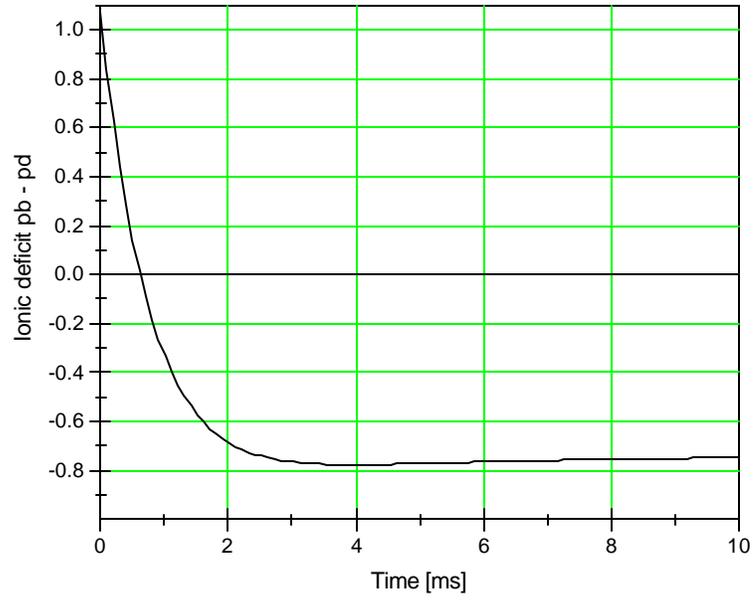

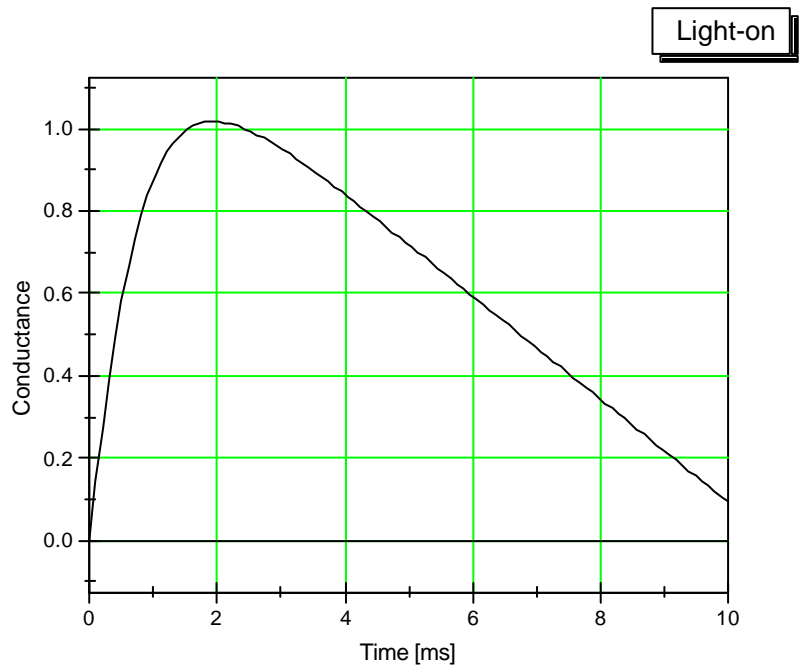

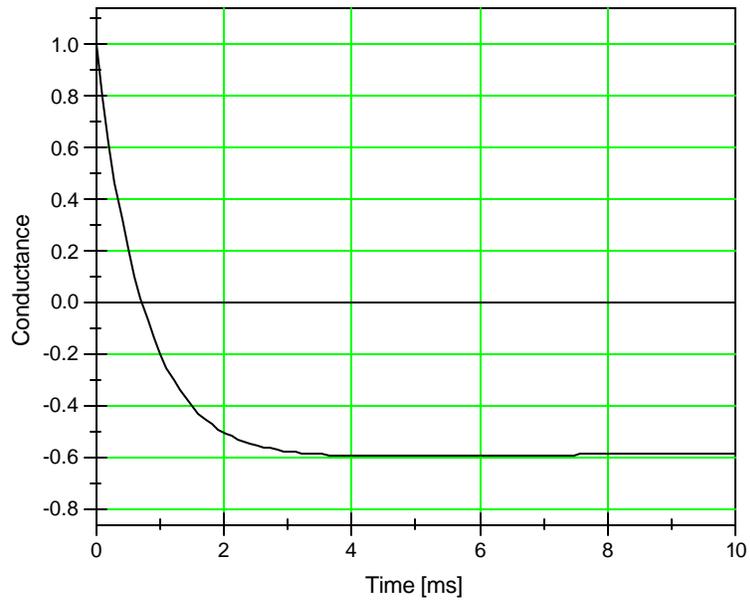

Figure 2. Calculated short-circuit photocarrier densities using the exact solutions of the rate equations listed in Table I for the slow-recombination negative-U mode. Both the solutions under light-on and light-off are depicted in arbitrary units. In (a) and (b) we show the unihole densities $p(t)_{on}$ and $p(t)_{off}$, in (c) and (d) the ionic deficit $p_b(t)_{on} - p_d(t)_{on}$ and $p_b(t)_{off} - p_d(t)_{off}$, respectively. The deficit turns surplus at the later stages, apparently reflecting the underestimated recombination rate. The resulting photoconductivities $\sigma_{on}(t)$ and $\sigma_{off}(t)$ are plotted separately in (e) and (f). $\sigma_{on}$ is seen to develop initially rapidly, then peak at the early stages of excitation, and decline more slowly to vanishing and assuming negative values beyond. The negative $\sigma_{on}$ obviously reflects the increased flow of readsorbing ions at the later stages. All the main features of the calculated photoconductivities may be seen in concert with the observed photovoltage transients in Figure 1 (d). Typical numerical parameters used for these and related calculations: $t_p = 3$ ms (LED) or $t_p = 10$ ms (Xe), $t_s = \sqrt{(G\gamma)^{-1}} = 1$ ms, $p_s = \sqrt{(G/\gamma)} = 10^{11} cm^{-3}$, $\gamma = 10^{-8} cm^3 s^{-1}$, $G = 10^{14}$ cm$^{-3}$s$^{-1}$ ($\eta = 0.1$, $c_\phi = 10^{-16}$ cm$^2$, $I = 10^9$ cm$^{-2}$s$^{-1}$, $N = 10^{22}$ cm$^{-3}$), $n = 10^{11}$ cm$^{-3}$, $\tau_b = 1$ ms, $\tau_d = 3$ ms, $\tau_s = 1$ ms, $\tau_r = \infty$ (slow) or $\tau_r = 0.1$ ms (fast).